\documentclass[twocolumn]{webofc}
\usepackage[varg]{txfonts}   
\usepackage{graphicx}
\usepackage{amssymb}
\usepackage{amsmath}
\usepackage{amsfonts}
\usepackage{dcolumn}
\usepackage{hyperref}
\usepackage{color}
\usepackage{comment}
\usepackage{breqn}
\usepackage{multicol}
\usepackage{multirow}
\usepackage[11pt]{moresize} 
\usepackage{flafter}
\def\costev{\cos(\theta_{\mathrm{e}\nu})}
\def\Ee{E_{\mathrm{e}}}
\def\pe{p_{\mathrm{e}}}
\def\pp{p_{\mathrm{p}}}
\def\tp{t_{\mathrm{p}}}
\def\me{m_{\mathrm{e}}}

\def\Vud{V_{\mathrm{ud}}}
\def\gA{g_{\mathrm{A}}}
\def\gV{g_{\mathrm{V}}}
\def\taun{\tau_{\mathrm{n}}}
\def\GF{G_{\mathrm{F}}}

\newdimen\arrayruleHwidth
\makeatletter
\def\Hline{\noalign{\ifnum0=`}\fi\hrule \@height \arrayruleHwidth
    \futurelet \@tempa\@xhline}
\def\HlineB{\noalign{\ifnum0=`}\fi\hrule \@height \arrayruleHwidth
    \futurelet \@tempa\@xhline}
\def\HlineM{\noalign{\ifnum0=`}\fi\color{Maroon}\hrule \@height
  \arrayruleHwidth \futurelet \@tempa\@xhline}
\def\hlineB{\noalign{\ifnum0=`}\fi\hrule \@height \arrayrulewidth
    \futurelet \@tempa\@xhline}
\def\hlineM{\noalign{\ifnum0=`}\fi\color{Maroon}\hrule \@height \arrayrulewidth
    \futurelet \@tempa\@xhline}
\makeatother

\newcommand{\td}{\text{d}}

\def \SCIe#1#2{\hbox{$#1\times 10^{#2}$}}

\begin{document}
\title{The Nab Experiment: A Precision Measurement of Unpolarized Neutron Beta Decay
}
%
%

\author{\firstname{J.} \lastname{Fry}\inst{1}\fnsep\thanks{\email{jaf2qc@virginia.edu}} \and
        \firstname{R.} \lastname{Alarcon}\inst{2}\and
        \firstname{S.} \lastname{Bae{\ss}ler}\inst{1,3}\and
        \firstname{S.} \lastname{Balascuta}\inst{2}\and
        \firstname{L.} \lastname{Barr\'{o}n Palos}\inst{4}\and
        \firstname{T.} \lastname{Bailey}\inst{5}\and
        \firstname{K.} \lastname{Bass}\inst{6}\and
        \firstname{N.} \lastname{Birge}\inst{6}\and
        \firstname{A.} \lastname{Blose}\inst{7}\and
        \firstname{D.} \lastname{Borissenko}\inst{1}\and
        \firstname{J. D.} \lastname{Bowman}\inst{3}\and
        \firstname{L. J.} \lastname{Broussard}\inst{3}\and
        \firstname{A. T.} \lastname{Bryant}\inst{1}\and
        \firstname{J.} \lastname{Byrne}\inst{8}\and
	\firstname{J. R.} \lastname{Calarco}\inst{3,6}\and
	\firstname{J.} \lastname{Caylor}\inst{6}\and
	\firstname{K.} \lastname{Chang}\inst{1}\and
	\firstname{T.} \lastname{Chupp}\inst{9}\and
	\firstname{T. V.} \lastname{Cianciolo}\inst{3}\and
	\firstname{C.} \lastname{Crawford}\inst{7}\and
	\firstname{X.} \lastname{Ding}\inst{1}\and
	\firstname{M.} \lastname{Doyle}\inst{1}\and
	\firstname{W.} \lastname{Fan}\inst{1}\and
	\firstname{W.} \lastname{Farrar}\inst{1}\and
	\firstname{N.} \lastname{Fomin}\inst{6}\and
	\firstname{E.} \lastname{Frle\v{z}}\inst{1}\and
	\firstname{M. T.} \lastname{Gericke}\inst{10}\and
	\firstname{M.} \lastname{Gervais}\inst{7}\and
	\firstname{F.} \lastname{Gl\"uck}\inst{11}\and
	\firstname{G. L.} \lastname{Greene}\inst{3,6}\and
	\firstname{R. K.} \lastname{Grzywacz}\inst{6}\and
	\firstname{V.} \lastname{Gudkov}\inst{12}\and
	\firstname{J.} \lastname{Hamblen}\inst{13}\and
	\firstname{C.} \lastname{Hayes}\inst{14}\and
	\firstname{C.} \lastname{Hendrus}\inst{9}\and
	\firstname{T.} \lastname{Ito}\inst{15}\and
	\firstname{A.} \lastname{Jezghani}\inst{7}\and
	\firstname{H.} \lastname{Li}\inst{1}\and
	\firstname{M.} \lastname{Makela}\inst{15}\and
	\firstname{N.} \lastname{Macsai}\inst{10}\and
	\firstname{J.} \lastname{Mammei}\inst{10}\and
	\firstname{R.} \lastname{Mammei}\inst{16}\and
	\firstname{M.} \lastname{Martinez}\inst{2}\and
	\firstname{D. G.} \lastname{Mathews}\inst{7}\and
	\firstname{M.} \lastname{McCrea}\inst{7}\and
	\firstname{P.} \lastname{McGaughey}\inst{15}\and
	\firstname{C. D.} \lastname{McLaughlin}\inst{1}\and
	\firstname{P.} \lastname{Mueller}\inst{3}\and
	\firstname{D.} \lastname{van Petten}\inst{1}\and
	\firstname{S. I.} \lastname{Penttil\"a}\inst{3}\and
	\firstname{D. E.} \lastname{Perryman}\inst{6}\and
	\firstname{R.} \lastname{Picker}\inst{17}\and
	\firstname{J.} \lastname{Pierce}\inst{3}\and
	\firstname{D.} \lastname{Po\v{c}ani\'c}\inst{1}\and
	\firstname{Y.} \lastname{Qian}\inst{1}\and
	\firstname{J.} \lastname{Ramsey}\inst{3}\and
	\firstname{G.} \lastname{Randall}\inst{2}\and
	\firstname{G.} \lastname{Riley}\inst{6}\and
	\firstname{K. P.} \lastname{Rykaczewski}\inst{3}\and
	\firstname{A.} \lastname{Salas-Bacci}\inst{1}\and
	\firstname{S.} \lastname{Samiei}\inst{1}\and
	\firstname{E. M.} \lastname{Scott}\inst{6}\and
	\firstname{T.} \lastname{Shelton}\inst{7}\and
	\firstname{S. K.} \lastname{Sjue}\inst{15}\and
	\firstname{A.} \lastname{Smith}\inst{1}\and
	\firstname{E.} \lastname{Smith}\inst{15}\and
	\firstname{E.} \lastname{Stevens}\inst{1}\and
	\firstname{J.} \lastname{Wexler}\inst{14}\and
	\firstname{R.} \lastname{Whitehead}\inst{6}\and
	\firstname{W. S.} \lastname{Wilburn}\inst{15}\and
	\firstname{A.} \lastname{Young}\inst{14}\and
	\firstname{B.} \lastname{Zeck}\inst{14}
}

\institute{Department of Physics, University of Virginia, Charlottesville, VA 22904-4714
\and
           Department of Physics, Arizona State University, Tempe, AZ 85287-1504
\and
           Physics Division, Oak Ridge National Laboratory, Oak Ridge, TN 37831
\and Universidad Nacional Aut\'{o}noma de M\'{e}xico, M\'{e}xico, D.F. 04510, M\'{e}xico
\and Department of Physics, North Carolina State University, Raleigh, NC 27695-8202 
\and Department of Physics and Astronomy, University of Tennessee, Knoxville, TN 37996
\and Department of Physics and Astronomy, University of Kentucky, Lexington, KY 40506
\and Department of Physics and Astronomy, University of Sussex, Brighton BN19RH, UK
\and University of Michigan, Ann Arbor, MI 48109
\and Department of Physics, University of Manitoba, Winnipeg, Manitoba, R3T 2N2, Canada
\and KIT, Universit\"at Karlsruhe (TH), Kaiserstra{\ss}e 12, 76131 Karlsruhe, Germany
\and Department of Physics and Astronomy, University of South Carolina, Columbia, SC 29208
\and Department of Chemistry and Physics, University of Tennessee at Chattanooga, Chattanooga, TN 37403
\and Department of Physics, North Carolina State University, Raleigh, NC 27695-8202
\and Los Alamos National Laboratory, Los Alamos, NM 87545
\and Department of Physics, University of Winnipeg, Winnipeg, Manitoba R3B2E9, Canada
\and TRIUMF, Vancouver, Canada, V6T 2A3
}

\abstract{%
  Neutron beta decay is one of the most fundamental processes in nuclear physics and provides sensitive means to uncover the details of the weak interaction. Neutron beta decay can evaluate the ratio of axial-vector to vector coupling constants in the standard model, $\lambda = g_A / g_V$, through multiple decay correlations. The Nab experiment will carry out measurements of the electron-neutrino correlation parameter $a$ with a precision of $\delta a / a = 10^{-3}$ and the Fierz interference term $b$ to $\delta b = 3\times10^{-3}$ in unpolarized free neutron beta decay. These results, along with a more precise measurement of the neutron lifetime, aim to deliver an independent determination of the ratio $\lambda$ with a precision of $\delta \lambda / \lambda = 0.03\%$ that will allow an evaluation of $V_{ud}$ and sensitively test CKM unitarity, independent of nuclear models. Nab utilizes a novel, long asymmetric spectrometer that guides the decay electron and proton to two large area silicon detectors in order to precisely determine the electron energy and an estimation of the proton momentum from the proton time of flight. The Nab spectrometer is being commissioned at the Fundamental Neutron Physics Beamline at the Spallation Neutron Source at Oak Ridge National Lab. We present an overview of the Nab experiment and recent updates on the spectrometer, analysis, and systematic effects.  
}
\maketitle
\section{Introduction and Motivation}
\label{intro}

Free neutron decay is one of the most fundamental and simplest weak interaction processes and serves as an illuminating tool to test our understanding of the Standard Model (SM). Much theoretical work has been done on neutron beta decay and its sensitivity to physics beyond the SM~\cite{GONZALEZALONSO}. To leading order, Jackson {\it et al.}~\cite{Jackson} describes the differential neutron decay rate parametrized by correlation coefficients $a$, $b$, $A$, $B$, $D$ etc. as

\begin{multline}
   \frac{dw}{dE_{\mathrm{e}} d\Omega_{\mathrm{e}} d\Omega_\nu} \propto \pe\Ee(E_0-\Ee)^2 \xi \ \\[1ex]
     \times \bigg[ 1
        + {a}\frac{\vec{p}_{\mathrm{e}}\cdot\vec{p}_\nu}{\Ee E_\nu}
        + {b}\frac{\me}{\Ee}
     + \langle \vec{\sigma}_{\mathrm{n}}\rangle\cdot
        \left({A}\frac{\vec{p}_{\mathrm{e}}}{\Ee}
          + {B}\frac{\vec{p}_\nu}{E_\nu}
	+ {D}\frac{\vec{p}_{\mathrm{e}}\times\vec{p}_\nu}{\Ee E_\nu }
           \right) \bigg]
           \label{fulldecayrate}
  \end{multline}
  
\noindent where $\pe$, $p_{\nu}$, $\Ee$, and $E_{\nu}$ are the momenta and energy of the decay electron and neutrino and $E_0$ is the endpoint energy of the electron spectrum. In the SM, $\xi$ = $\GF^2\Vud^2(1+3|\lambda|^2)$, where $\GF$ is the Fermi constant, $\Vud$ is the first diagonal term in the CKM matrix, and $\lambda$ is the ratio of the axial vector to vector coupling constants, $\lambda = \gA/\gV$. Lastly, $\sigma_{\mathrm{n}}$ is the neutron spin and the correlation coefficients $a$, $b$, $A$, $B$, and $D$ are to be determined from experiment. In the Nab experiment, we study unpolarized neutron beta decay, which can access the coefficients $a$ and $b$. All the correlation coefficients other than $b$ depend on $\lambda$, specifically the electron-neutrino coefficient $a=(1-|\lambda|^2)/(1+3|\lambda|^2)$. In the SM, the Fierz interference term is defined as $b$ = 0 and a non-zero determination of $b$ is sensitive to scalar and tensor non-SM processes, competitive with muon decay and LHC~\cite{Baessler2014}. 

The total neutron decay rate $w$ or the neutron lifetime $\taun$, depends on $\Vud$ and $\lambda$, as $w = 1/\taun \propto |\Vud|^2 \GF^2 (1 + 3 |\lambda|^2)$. Over the last few years, efforts of the UCNA and PERKEO II groups have measured the beta asymmetry $A={-2|\lambda|(|\lambda|+1)}/(1+3|\lambda|^2)$ and found $\lambda$ = -1.2772(20) and -1.2761($^{+14}_{-17}$), respectively~\cite{UCNA2017,PERKEOII}, which is in some tension with the previous experiments~\cite{nPDG}. The PERKEO III experiment announced preliminary results at this conference with an error of $\Delta \lambda / \lambda$ = 0.06\%, and the final results will be published soon~\cite{PPNS2018}. The future polarized experiments PERC~\cite{PERC} and UCNA+ (an upgrade of~\cite{UCNA2017}), as well as a polarized version of Nab, aim to improve the precision of $A$. Independent extractions of $\lambda$ from different correlation coefficients offer a different set of systematic uncertainties and consistency checks and are necessary to entangle $\Vud$ from the neutron lifetime. Measurements of the neutron lifetime in the beam~\cite{YueBeam} and bottle method~\cite{SEREBROV2005,PICHLMAIER2010,MamboIReanalysis,ARZUMANOV2015,Pattie2017,Serebrov2017} produce a 3$\sigma$ discrepancy. Both methods are pursuing higher precision measurements to resolve this discrepancy. Additionally, the recent updated universal radiative correction $\Delta_R^V$~\cite{Seng} shifts $\Vud$ extracted from $0^+\to0^+$ decays~\cite{Hardy} downward from 0.97417(21) to 0.97366(15), a 4$\sigma$ deviation from CKM unitarity~\cite{CKMPDG}. Extracting $\Vud$ from the neutron lifetime and neutron beta decay correlations is important as neutron beta decay carries no nuclear structure uncertainties. The neutron sector must measure the neutron lifetime to $\sim0.3$\,s and $\lambda$ to $\sim3\times10^{-4}$ to competitively test the most precise determination of $\Vud$ from $0^+\to0^+$ decays~\cite{Hardy}. 

The Nab experiment aims for a high precision measurement of $a$ with an expected error of $\Delta a / a \sim 1 \times 10^{-3}$ or $\Delta \lambda / \lambda \sim 3 \times 10^{-4}$, about a factor of 40 more precise than the most precise extractions to date ~\cite{Stratowa,Byrne2002,aCORN} and a factor of 9 more precise than the preliminary results of the aSPECT experiment announced at this conference~\cite{PPNS2018}. 

\begin{figure}[t]
\centering
\includegraphics[width=1\linewidth,keepaspectratio]{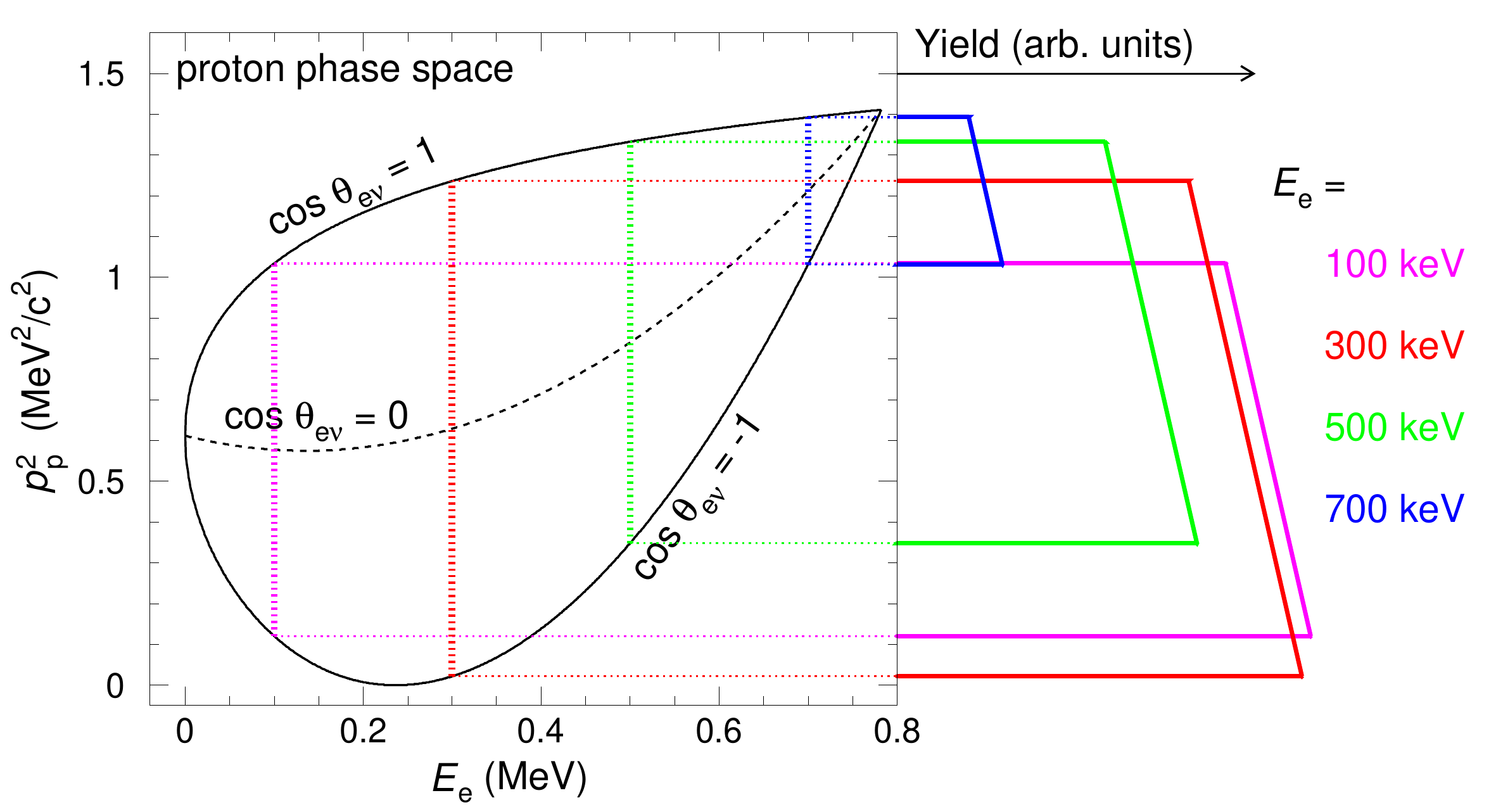}
\caption{Plot of the phase space of $\costev$ as a function of $\pp^2$ and $\Ee$ with projections of the squared proton momentum distribution $P(\pp^2)$ for a fixed electron energy $\Ee$ and an ideal detection system of $\pp^2 \propto$ 1/$\tp^2$ (denoted trapeziums in the text). }
\label{phasespace}       
\end{figure}

\section{Measurement Principles}

The electron-neutrino correlation coefficient $a$ requires an extraction of the opening angle between the electron and neutrino, $\costev$. If we consider the relativistic kinematics, conservation of momentum yields

\begin{equation}
\costev = \frac{\pp^2 - \pe^2 - p_{\nu}^2}{2\pe p_{\nu}}.
\label{costhetaev}
\end{equation}

\noindent When radiative corrections and recoil corrections are neglected, $\costev$ is linearly related to $\pp^2$ for a fixed $\Ee$ since the neutrino energy can be related to the electron energy ($\Ee + E_{\nu} = E_0$). Thus a measurement of $\pp^2$ and $\Ee$ can determine $\costev$. Figure \ref{phasespace} shows allowed values of $\pp^2$ and $\Ee$ from the phase space of neutron beta decay. If we assume $b$ = 0 in the SM and $\langle \vec{\sigma}_{\mathrm{n}}\rangle$ = 0, equation~\ref{fulldecayrate} simplifies to $dw \propto 1 + a \beta \costev$, where $a$ is the electron-neutrino correlation coefficient in question and $\beta$ = $v_{\mathrm{e}}/c$. For a fixed $\Ee$, the decay rate will have a slope of $a$ in the distribution of $\pp^2$ as shown in figure~\ref{phasespace}. The fact that a value of $a$ can be extracted for each electron energy gives consistency checks for systematic effects that depend on electron energy. The Fierz interference term $b$ is measured both simultaneously in the Nab-$a$ configuration (explained below) and in a separate Nab-$b$ configuration through the shape of the electron energy spectrum. The remainder of this paper focuses on the extraction of $a$.

\begin{figure}[t]
\centering
\includegraphics[width=.9\linewidth,keepaspectratio]{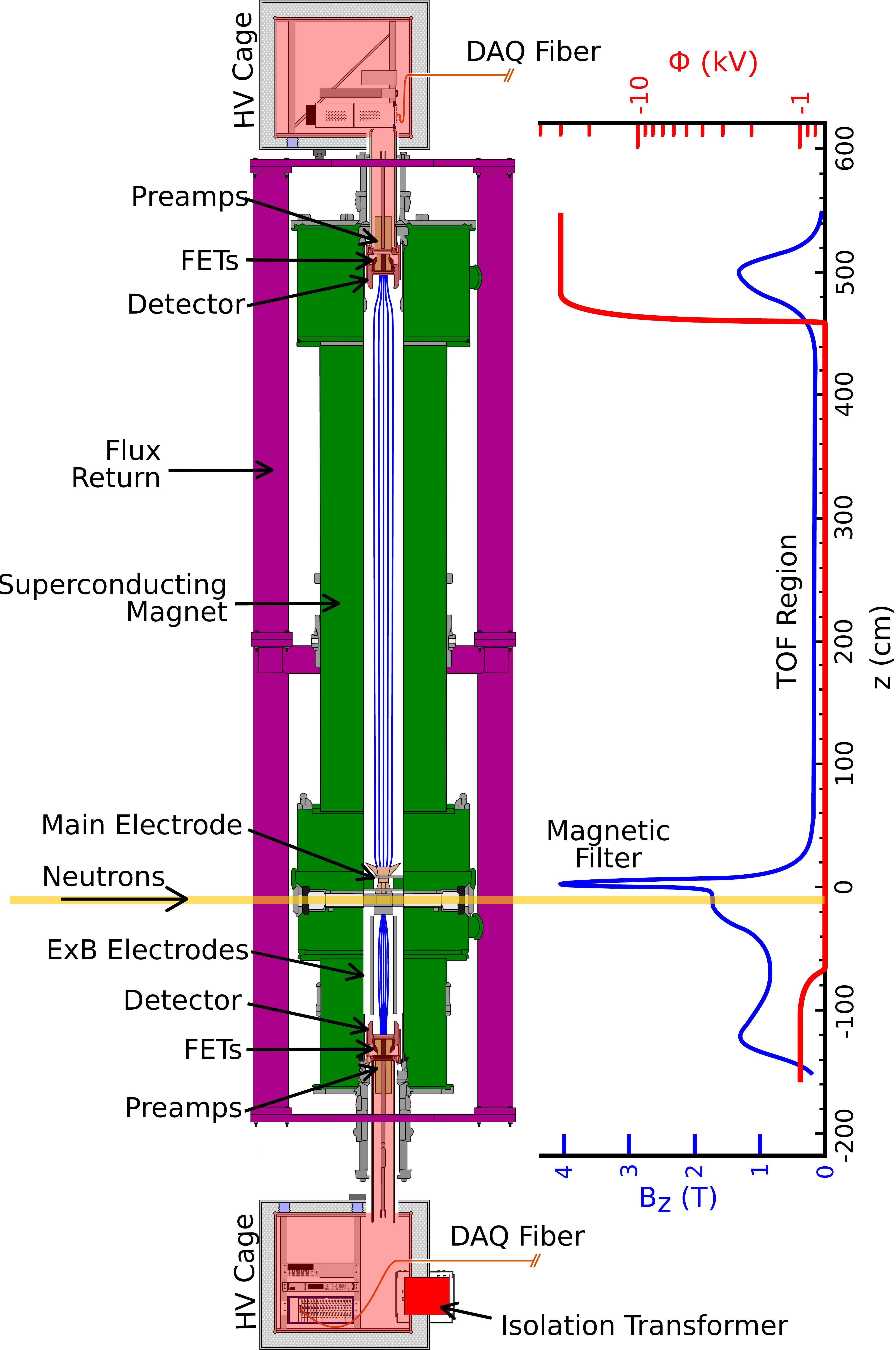}
\caption{A diagram of the Nab spectrometer with details of the electronics, detector system, electrodes, and electric and magnetic field profiles.}
\label{spec}       
\end{figure}

To extract $a$, the proton momentum $\pp$ and electron energy $\Ee$ must be determined. Since the endpoint of the proton energy spectrum from neutron beta decay is 751\,eV, a direct determination of its momentum is difficult. Thus, we use a long, asymmetric magnetic spectrometer to estimate $\pp$ from the time of flight (TOF) of the proton, $\tp$, and reconstruct $\Ee$ from energy deposited in Si detectors. We will apply a Monte-Carlo correction for the small electron TOF and this effect is addressed in the systematics table. Figure \ref{spec} shows the Nab magnetic spectrometer with details of the magnetic and electric field profiles. Note that $z$ = 0 is defined as the magnetic filter peak and the center of the decay volume is $z$ = -13.2\,cm. The electrons and protons produced from neutron beta decay spiral along the field lines and are guided to detectors at each end of the spectrometer, $\sim$-1.1\,m below and $\sim$5.1\,m above the decay volume.

Both detectors are placed in a 1.3\,T field and an accelerating potential of -30\,kV and -1\,kV is maintained at the top and bottom detectors, respectively, by cylindrical electrodes. For the Nab-$a$ configuration, the accelerating potential in the upper detector is required for the protons to be detected, while electrons can be detected in either detector. The -1\,kV potential on the bottom detector and the $E\times B$ electrodes between the decay volume and lower detector prevent protons from being reflected from the lower to upper detector. The Si detectors were developed for Nab by Micron Semiconductor Ltd [20] to detect electrons and protons (with the accelerating potential). The detectors are segmented into 127 pixels for position determination, with observed energy resolution of 3 keV (FWHM) and a 40 ns rise time (10\%-90\% amplitude). Detectors will undergo full system testing at LANL and the University of Manitoba in the next few months. Electron backscattering is mitigated as the magnetic field lines will always guide bouncing electrons to one of the detectors. Electron backscattering and energy reconstruction has been studied within the collaboration, and remains an important topic of continued study. The initial characterization of the detectors at the Triangle Universities Nuclear Laboratory (TUNL) proton accelerator showed a dead layer of 100\,nm and a resolution near 3\,kV ~\cite{salasNIM,BROUSSARD2017}, meeting the needs of the Nab experiment. Another recent update of the Nab detectors and electronics can be found here~\cite{Leah}. 

\section{Details of the Nab Spectrometer and Proton Time of Flight}

Neutrons from the FnPB beamline~\cite{FOMIN} at the SNS at ORNL pass through the spectrometer decay volume. Electrons and protons from neutron beta decay are born in a magnetic field of 1.7\,T. Above the decay volume, a strong magnetic curvature (with peak field of 4\,T) acts as a magnetic filter to accept only protons with momentum within a narrow upward cone along the spectrometer axis, creating a minimum accepted angle $\theta_{0,min}$. Subsequently, the field expansion from the magnetic filter to the long TOF region (0.2\,T) largely longitudinalizes the momentum and adiabatically guides the charged particles to the upper detector. Figure \ref{spec} shows a diagram of the Nab spectrometer and the electric and magnetic field profiles. The shortest TOF for an upward proton is about 13\,$\mu$s and the shortest TOF for a downward electron is about 5\,ns. At 1.4\,MW SNS primary beam power, we estimate 1600\,decays/s, equivalent to 200\,protons/s in the upper detector. Nab plans to collect several samples of $10^9$ coincidence events in several runs over $\sim$2 years running cycle at the SNS to accomplish the statistical demands of the experiment. The magnet is now installed on the FnPB beamline and commissioning of the magnet and subsystems is ongoing.

Nab will make an estimate of $p_p$ from $t_p$. The relationship between $p_p$ and $t_p$ depends on the guiding center of the field lines, electrostatic potential experienced in the spectrometer, the unobserved angle between the born momentum and magnetic field vectors, and the size of the neutron beam in the decay volume, as well as other smaller systematic effects. 
For an adiabatically expanding field, $t_p$ is given by an integral along the guiding center: 

\begin{equation}
   t_p
     = \frac{m_p}{p_p} \int_{z_0}^l \frac{\,dz}
      {\sqrt{1-{\frac{B(z)}{B_0}\sin^2\theta_{\text{0}}} + {\frac{q(V(z)-V_0)}{E_{0}}}}}
      \label{MethodA}
\end{equation}

\begin{figure*}[t]
\centering
\centering
\includegraphics[width=1\linewidth,keepaspectratio]{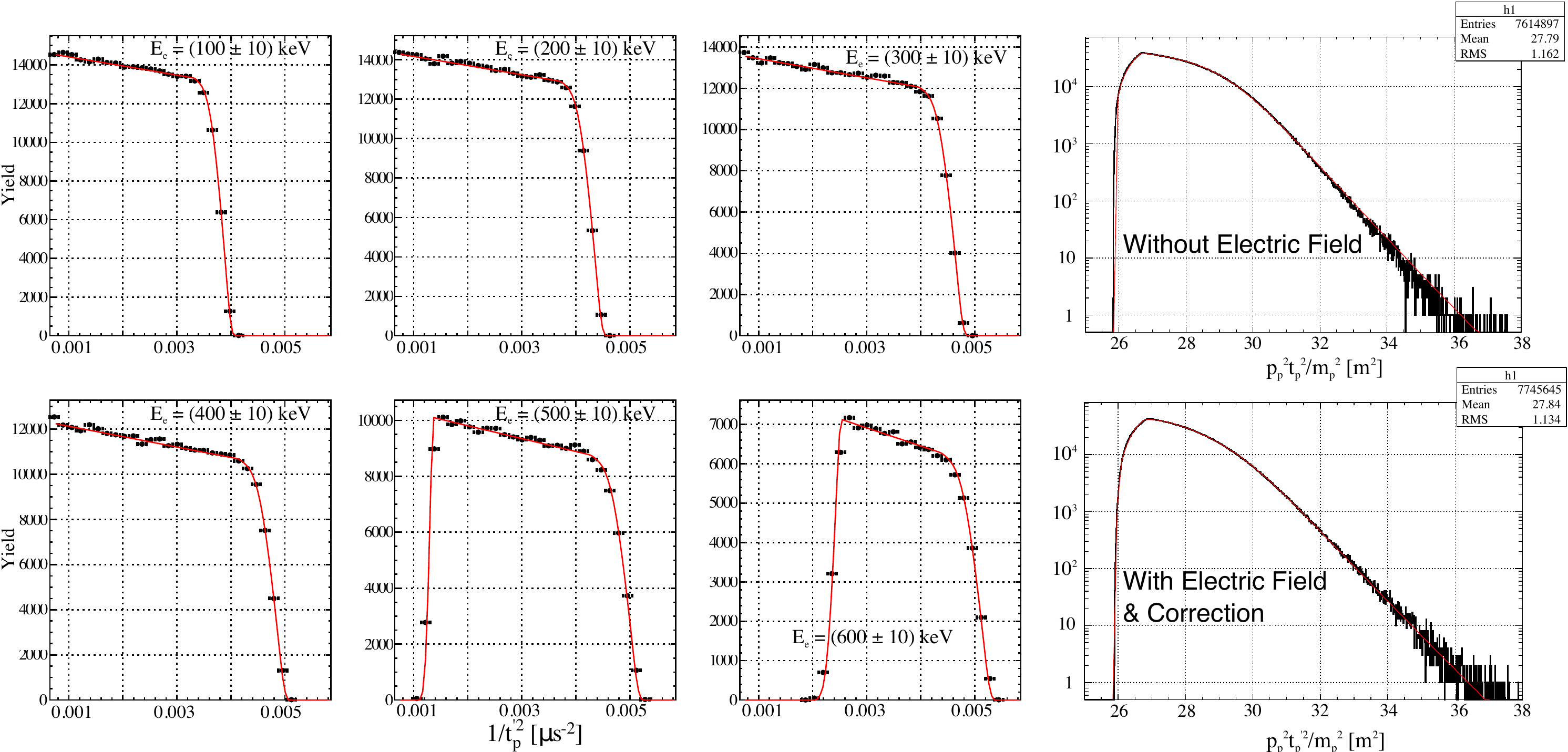}
\caption{( Left) Simulated $1/t_p^{'2}$ distributions for different electron energies (for $a = -0.103$). The red lines are fits to the $1/t_p^{'2}$ which take into account the detector response function through analysis `Method A' as described in the text. (Right) Simulated detector response function for without electric field ($p_p^2\cdot t_p^2/m_p^2$) and with electric field and corrections ($p_p^2\cdot t_p^{'2}/m_p^2$).}
\label{trapsFit}       
\end{figure*}

\noindent where $z_0$, $\theta_0$, $B_0$, $V_0$, $p_p$, and $E_0$ are the decay coordinate, angle of the proton momentum with respect to the magnetic field vector, magnetic field magnitude, electric potential, and magnitudes of the momentum and energy at birth, respectively. $B(z)$ and $V(z)$ are the magnetic field and potential as a function of $z$, and $l$ is the length from the center of the decay volume to the upper detector. The unobserved quantities $z_0$ and $\theta_0$ lead to imperfect knowledge of the reconstruction of $p_p^2$ from $1/t_p^2$. These properties form what we call the spectrometer response function of the $1/t_p^2$ distribution. An ideal spectrometer response is a one-to-one delta function between $p_p$ and $t_p$, but the actual spectrometer response will have nonzero width in $1/t_p^2$ for a fixed $p_p^2$. This results in a smearing of the edges in the $1/t_p^2$ distributions (trapeziums, as shown in figure~\ref{trapsFit}). The analysis strategies for Nab need to understand or parameterize the spectrometer response function to extract a reliable value of $a$.

To fully understand the spectrometer response and other systematic effects, a detailed Monte Carlo simulation in Geant4 has been written and benchmarked. Charged particles from neutron decay are stepped through magnetic and electric fields using the Geant4 Cash-Karp 4/5$^{th}$-order Runge-Kutta-Fehlberg method~\cite{Geant4}. The magnetic and electric fields are analytically determined using the method and code in reference~\cite{Ferenc} and a 1D Radial Series Expansion (RSE)~\cite{RSE} is used to expand the field into cylindrically symmetric 2D coordinates to speed up runtime (see equation \ref{RSEeq} below). The final kinematics of the protons and electrons at the detector are stored for further analysis.

We use these realistic simulation data to test our analysis algorithms, which are described in~\cite{NabProp,Baessler2014}. One method, called `Method A' used to treat the integral in equation~\ref{MethodA}, utilizes Monte Carlo simulations with and without electric field to find a mapping between the two so that distributions in a simulation with electric field can be corrected for the electrostatic term. Figure~\ref{trapsFit} (right) shows the detector response function, $p_p^2\cdot t_p^2/m_p^2$, distributions for no electric field and with electric field and corrections. We find the mapping gives sufficient precision to carry out the analysis procedure. We denote primed variables such as $t_p^{\prime}$ as the electrostatic corrected variables using such a mapping. This approximation eliminates the electrostatic term and leaves us with the second term in the integral containing the magnetic field and $\sin(\theta_0)^2$. For small angles, this can be expanded into a Taylor series expansion including an additional term that is needed for particles with $\theta_0$ close to the critical angle $\theta_{0,min}$:

\begin{align}
    { p_{p}}&=\frac{m_{p}}{{ t_{p}'}} \int{\frac{dz}{\sqrt{1 - {\frac{B(z)}{B_{0}}sin^{2}(\theta_{0})}}}} \\
    &\nonumber
    =\frac{m_{p}}{t_{p}'} \left( L - \eta_A \text{ln} \frac{\cos(\theta_{0})-{\cos(\theta_{0,min})}}{1-{\cos(\theta_{0,min})}} + \alpha_A(1-\cos(\theta_{0})) \right.\\
    &\nonumber
    + \left. \beta_A(1-cos(\theta_{0}))^{2} + \gamma_A(1-cos(\theta_{0}))^{3} \right).
    \end{align}

\noindent Here, $L$ is the effective length of the spectrometer, the $\eta$ term is an analytic expression for the TOF through the filter, and $\alpha_A$, $\beta_A$, and $\gamma_A$ are Taylor series expansion coefficients in $(1-\cos(\theta_0))$. To obtain the parameters $\eta_A$, $\alpha_A$, $\beta_A$, and $\gamma_A$, we either fit the simulated data of $p_p\cdot t_p^'/m_p$, fit the edges of the $1/t_p^{'2}$ distributions~\cite{Baessler2014}, or a combination of both. Then the $1/t_p^{'2}$ distributions are fit to extract $a$. Figure~\ref{trapsFit} (left) shows the simulated $1/t_p^{'2}$ distributions for different energy slices as well as the fit results using the method described above (in red). 

\subsection{The Magnetic Field of the Spectrometer and Associated Systematics}

\begin{figure*}[t]
\centering
\centering
\includegraphics[width=1\linewidth,keepaspectratio]{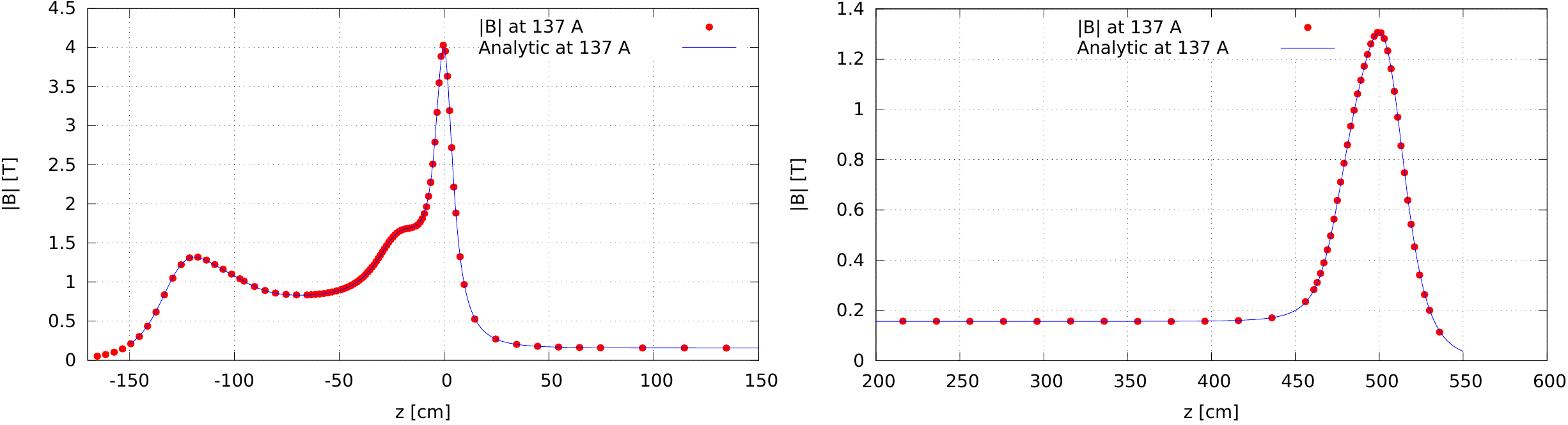}
\caption{Magnetic field on-axis measured during acceptance testing at the SNS (data in red, prediction in blue). The result agrees with prediction at the level of precision of this preliminary measurement.}
\label{Acceptance}       
\end{figure*}

Details of the magnetic field dependence on $z$ play an important role in the extraction of $a$. For systematic uncertainties,
      $\gamma$ = $-(1/B)(\text{d}^2 B/\text{d}z^2)$, the field curvature in the filter, and ratios
      $r_B$ = $B_{\text{TOF}}/B_0$ and $r_{B,DV}$ = $B_{\text{DV}}/B_0$, of the 
      magnetic field in the time-of-flight section (TOF) and the decay
      volume (DV), respectively, to the filter peak (0), address the largest systematics uncertainties~\cite{NabProp}.
      The required sensitivity is:


\begin{equation}
\frac{\Delta r_{B}}{r_{B}} =  10^{-2}, \hspace{0.7pc}\frac{\Delta r_{B,DV}}{r_{B,DV}} = 10^{-2}, \hspace{0.7pc}\frac{\Delta \gamma}{\gamma} < 2 \times 10^{-2}
\end{equation}

In addition, the field needs to be known everywhere to a similar precision for input into detailed Geant4 simulations. A detailed on-axis measurement, followed by an off-axis measurement will be performed in the next few months using a Group3 Hall probe calibrated to better than $10^{-3}$. We use the on-axis measurement to expand the field off-axis and use the off-axis measurement as a consistency check. We use two analyses to expand the field off-axis: an RSE and a modified Bessel function expansion (MBFE) approach. Cylindrical symmetry is required for these analyses and we expect this symmetry to a large degree. The principle is the same for each method -- use data from a non-equispaced grid on-axis, $B(r=0,z)$ and expand off-axis to obtain $B(r,z)$. The RSE originates from work in understanding axisymmetric fields for electron transport~\cite{RSE}, among other applications. For an axisymmetric configuration of coils, an off-axis expansion can be performed using only the on-axis vertical field $B_z(r=0,z)\equiv B_{0,z}$ and its derivatives ${\td^n B_{0,z}}/{\td z^n}$. We have found that including up to the 6$^{th}$ derivative satisfies the required precision. The general expansion is:

\begin{equation}
\begin{split}
B_z(r,z) =& \sum\limits_{n=0}^{\infty} \frac{(-1)^n}{(2^nn!)^2}\frac{\td^{2n}B_{0,z}}{\td z^{2n}} r^{2n}\\
B_r(r,z) =& -\sum\limits_{n=0}^{\infty} \frac{(-1)^n}{n!(n+1)2^{2n+1}}\frac{\td^{2n+1}B_{0,z}}{\td z^{2n+1}} r^{2n+1}.
\label{RSEeq}
\end{split}
\end{equation}

The MBFE can be derived from a separable magnetic potential $\phi(r,z)  = \sum_k a_k e^{ikz} f(r)$. Solving Laplace's equation and taking the divergence, one obtains $f(r) = I_0(kr)$, the modified Bessel's function. Letting $k = 2 \pi j/L$, we have:

\begin{equation}
\begin{split}
B_z(r,z) =& \sum_j i a_j \left(\frac{2 \pi j}{L}\right) e^{\frac{i 2\pi j z}{L}} I_0 \left(\frac{2 \pi j r}{L}\right) \\
B_r(r,z) =& \sum_j a_j \left(\frac{2 \pi j}{L}\right) e^{\frac{i 2\pi j z}{L}} I_1 \left(\frac{2 \pi j r}{L}\right).
\end{split}
\end{equation}

\noindent To obtain the $a_j$ coefficients, we use the on-axis field map of $B_z$ at $r=0$ since the modified Bessel function is zero. Then, the off-axis expansion for $B_z(r,z)$ and $B_r(r,z)$ is simply a multiplicative factor by the modified Bessel function.

The acceptance tests for the Nab spectrometer magnet were conducted at the SNS in March, 2018. During the tests, the first measurements of the field were taken on-axis and compared with the analytical prediction. The measurement procedure had an error of about 1-2\% due to the positioning of the probe. The results are shown in figure \ref{Acceptance} and agree well with prediction at the precision of this measurement. More detailed magnetic field mapping will be carried out to achieve the ultimate required precision.

The important systematics in Nab have been discussed in references~\cite{POCANIC2009,NabProp,Baessler2014}. Below in table~\ref{sys} is an updated list of the systematics from these references. Please see these references for more details. 

\begin{table*}[]
      \label{sys}
\centering
\caption{Updated table of systematics from references~\cite{POCANIC2009,NabProp,Baessler2014}. The (*) indicates a fit parameter. Small in the third column indicates $(\Delta a/a)_{\text{SYST}} < \SCIe{1}{-4}$.}
\begin{tabular}{lcc} 
      \HlineB \\[-5pt]
       Experimental parameter & Principal specification
                       {(comment)} &
                                    {($\Delta a/a)_{\text{SYST}}$} \\[2pt]
      \hlineB \\[-5pt]
      {Magnetic field:}  \\
       \quad curvature at pinch &
        $\Delta \gamma/\gamma=2$\% with
                 $\gamma=(\td^2B_z(z)/\td z^2)/B_z(0)$
                                          & \SCIe{5.3}{-4}\\
       \quad ratio $r_{\text{B}}=B_{\text{TOF}}/B_0$ &
                     $(\Delta r_B)/r_B = 1\%$  & \SCIe{2.2}{-4} \\
       \quad  ratio $r_{\text{B,DV}}=B_{\text{DV}}/B_0$ &
        $(\Delta r_{\text{B,DV}})/r_{\text{B,DV}} = 1\%$  & \SCIe{1.8}{-4} \\[2pt]
      \multicolumn{2}{l}{{$L_{\text{TOF}}$},
                                         length of TOF region} & (*) \\[2pt]
      {$U$ inhomogeneity}: \\
       \quad in decay / filter region &
          $|U_{\text{F}}-U_{\text{DV}}|\, < 10$\,mV & \SCIe{5}{-4} \\
       \quad  in TOF region &
          $|U_{\text{F}}-U_{\text{TOF}}|\, < 200$\,mV & \SCIe{2.2}{-4} \\[2pt]
      {Neutron beam}: \\
       \quad position &
                   $\Delta \langle z_{\,\text{DV}}\rangle < 2$\,mm
                      & \SCIe{1.7}{-4} \\
       \quad profile (incl.\ edge effect) &
                          slope at edges $<$ 10\%/cm & \SCIe{2.5}{-4} \\
       \quad Doppler effect &  {(analytical correction)}
                                                             &  small \\
       \quad unwanted beam polarization &
              $\Delta\langle P_{\text{n}}\rangle < 2 \cdot 10^{-5}$
               {(with spin flipper)} &  $1\times 10^{-4}$ \\[3pt]
      \multicolumn{2}{l}{
        {Adiabaticity} of proton motion} & \SCIe{1}{-4}\\[2pt]
      {Detector effects}: \\
       \quad $E_{\text{e}}$ calibration &
         $\Delta E_{\text{e}} < 200$\,eV &  $2\cdot 10^{-4}$ \\
            \quad shape of $E_{\text{e}}$ response &
         $\Delta N_{\text{tail}}/N_{\text{tail}} < 1\%$ &\SCIe{4.4}{-4}  \\
       \quad proton trigger efficiency &
         $\epsilon_{\text{p}} < 100$\,ppm/keV & \SCIe{3.4}{-4} \\
       \quad TOF shift (det./electronics) &
         $\Delta t_{\text{p}} < 0.3$\,ns  &     \SCIe{3}{-4} \\[2pt]
      {TOF in accel. region}  & 
          $\Delta r_{\text{GROUND EL.}} < 0.5$\,mm {(preliminary)}
                                                  & \SCIe{3.4}{-4} \\[2pt] 
	{electron TOF}  & 
          (analytical correction)
                                                  & small \\[2pt]                                                   
      {BGD/accid.\ coinc's} 
         & {(will subtract out of time coinc)} & small \\[2pt]
      {Residual gas}  & 
          $p < 2 \cdot 10^{-9}$\,torr            & \SCIe{3.8}{-4} \\[2pt]
      \hlineB \\[-7pt]
      Overall sum & & \SCIe{1.2}{-3} \\
      \HlineB
      \label{sys}
    \end{tabular} \\[-5pt]
    \end{table*}

\normalsize
\section{Summary}
The Nab experiment aims for a measurement of $a$, the
    electron-neutrino correlation parameter, in neutron beta decay, with
    $\sim 10^{-3}$ relative precision.  This result will enable an
    independent precise determination of $\lambda = \gA/\gV$.  Once the
    neutron lifetime is measured with an uncertainty < 0.3\,s the
    expected Nab value of $\lambda$ will provide competitive precision
    to nuclear superallowed $0^+ \to 0^+$ decays in determining $V_{ud}$
    and testing CKM unitarity.
    

The magnet is installed on the FnPB at the SNS and initial tests of
     the magnetic field show that the results are consistent with expectations.
     Detailed measurements of the field will commence shortly.
     Installation of other beamline components is underway and  
     commissioning of the experiment will begin soon.  In parallel,
     simulation studies of systematics remain an ongoing area of intense
     work.
     
We acknowledge the support of the U.S. Department of Energy, the National Science Foundation, the University of Virginia, Arizona State University, and the Natural Sciences and Engineering Research Council of Canada.
     

%
 \bibliography{mybib}{}
\end{document}